# Growth of two-inch free-standing heteroepitaxial diamond on Ir/YSZ/Si (001) substrates via laser-patterned templates

Pengfei Qu[1,2], Peng Jin[1,2*], Guangdi Zhou[1,2], Zhen Wang[1,2], Zhanguo Wang[1,2]

1. Laboratory of Solid-State Optoelectronic Information Technology, Institute of Semiconductors, Chinese Academy of Sciences, Beijing 100083, P. R. China

2. Center of Materials Science and Optoelectronics Engineering, University of Chinese Academy of Sciences, Beijing 100049, P.R. China

* Corresponding author. Laboratory of Solid-State Optoelectronic Information Technology, Institute of Semiconductors, Chinese Academy of Sciences, Beijing 100083, P. R. China

E-mail address: pengjin@semi.ac.cn.

**Abstract**

In this paper, 2-inch free-standing diamonds were prepared by using heteroepitaxy on composite Ir/YSZ/Si (001) substrates. To release stress, patterned templates were fabricated using laser etching after the initial growth of 50-nm-diamond. Then, the subsequent growth was completed on a patterned template. The full width at half maximum of the diamond (400) and (311) X-ray rocking curves were 313.5 and 359.3 arcsecs, respectively. Strong band-edge emission in the cathodoluminescence spectrum of the resulting diamond revealed excellent crystalline quality. Furthermore, the 2D mapping of Raman spectra was conducted on a 2 mm × 2 mm area located at the center of the 2-inch sample with a thickness of 400 μm. The result showed an average peak width of 2.85 ± 0.36 cm$^{-1}$ and residual stress of -0.03 ± 0.37 GPa. The dislocation density, determined by counting etching pits generated from $H_2/O_2$ plasma etching, was estimated to be around 2.2 × 10$^7$ cm$^{-2}$. These results evidence that the laser-patterned method can effectively release stress during the growth of large-size diamonds, offering a simpler and more cost-effective alternative to the traditional photolithography-patterned scheme.

The interest in diamond is due to its amalgamation of remarkable physical properties, including a wide bandgap energy (~5.5 eV), a high breakdown field (10 MV·cm$^{-1}$),[1] high carrier mobility (4500 cm$^2$·V$^{-1}$·s$^{-1}$ for electrons and 3800 cm$^2$·V$^{-1}$·s$^{-1}$ for holes),[2] and high thermal conductivity (24 W·cm$^{-1}$·K$^{-1}$).[3] The last two decades have witnessed significant developments in the growth of high-quality single-crystal diamonds using the plasma-assisted chemical vapor deposition technique, driven by highly promising applications of diamonds in power electronics,[4] deep-ultraviolet and high-energy particle detectors,[5,6] as well as quantum devices based on color centers.[7]

A high-quality inch-sized monocrystalline wafer is fundamental for massive semiconductor device research. Nevertheless, preparing such diamond wafers constitutes one of the major technological challenges in commercializing diamond materials and devices. Over several decades of development, the heteroepitaxial growth of diamonds on Ir composite substrates has been recognized as an effective method for producing inch-sized monocrystalline diamonds.[8] This effectiveness primarily arises from the unique dissolution-precipitation mechanism between the Ir lattice and carbon atoms during the bias-enhanced nucleation (BEN) process. This mechanism enables achieving oriented nucleation densities on Ir surpassing $10^8$ cm$^{-2}$, far exceeding what can be attained on other hetero-substrates.

Heteroepitaxy theoretically promises to produce single-crystal diamonds as large as the underlying substrates. In well-studied Ir composite substrates, such as Ir/SrTiO$_3$/Si,[9,10] Ir/Sapphire,[11–14] Ir/MgO,[15–17] and Ir/YSZ/Si,[18–20] silicon and sapphire wafers reaching sizes up to 12 inches are commercially available. However, in the actual preparation process, achieving large-sized heteroepitaxial diamonds remains challenging. The disparity in lattice parameters between hetero-substrates and diamonds leads to high threading dislocation densities, typically in the range of $10^7$-$10^9$ cm$^{-2}$ for a few hundred micrometers thickness.[21] Furthermore, the inhomogeneous microwave plasma density and its spatial distribution over the surface of large-sized samples leads to noticeable temperature variations across the substrate. This non-uniform temperature distribution complicates strain control within the diamond-iridium composite system. Additionally, the thermal mismatch stemming from differences in thermal expansion coefficients induces stress of up to several *GPa* in diamond thin films, resulting in the cracking of diamond epitaxial layers during the cooling process.

Significant endeavors have been dedicated to diminishing dislocation densities and releasing stress in heteroepitaxial diamonds. In 2017, Schreck et al. fabricated a ~92-mm-diameter freestanding heteroepitaxial diamond plate with a thickness of 1.6 mm on Ir/YSZ/Si (001), the largest single-crystal diamond ever reported.[20] The diamond exhibited promising crystalline quality, as evidenced by a full-width at half maximum (FWHM) of 230 arcsecs for the X-ray rocking curve (XRC) of the diamond (400) and 432 arcsecs for (311), respectively. This achievement was facilitated by applying high-power conditions (915 MHz) in their microwave-plasma chemical vapor deposition (MPCVD) setup, expanding the plasma dimensions and thereby ensuring a more homogeneous growth environment for the diamonds. Besides, Aida et al. developed a novel microneedle method, introducing micro-patterns in the initial stages of heteroepitaxial diamond growth, to decrease the dislocation density and ease the strain in heteroepitaxy.[11,22,23] Kim et al. grew 1-inch high-quality self-supported diamonds with a thickness ranging from 500-600 μm after double-sided polishing on Ir (001)/sapphire (11$\bar{2}$0) using the microneedle method.[13] The FWHM of the XRCs are 113.4 and 234.0 arcsecs for the (400) and (311) crystallographic planes, respectively. Subsequently, they accomplished 2-inch high-quality freestanding monocrystalline diamonds on sapphire (11$\bar{2}$0) substrates with a misoriented angle of up to 7° in a step-flow growth mode.[14] This diamond has the lowest FWHM of 98.35 and 175.3 arcseconds for the XRC (400) and (311), respectively, and a dislocation density of around $10^7$ cm$^{-2}$.

This letter presents a practical and dependable approach for growing sizable diamond single crystals. We introduce the utilization of a laser beam to create patterns on the surface of a 50 nm diamond layer grown on BEN-treated Ir/YSZ/Si (001) substrates. These laser-patterned templates effectively ease the stress within the diamond layer, thereby facilitating the growth of a robust 2-inch freestanding diamond. The resultant diamond remains intact, devoid of cracks, and promising crystal quality and low dislocation density.

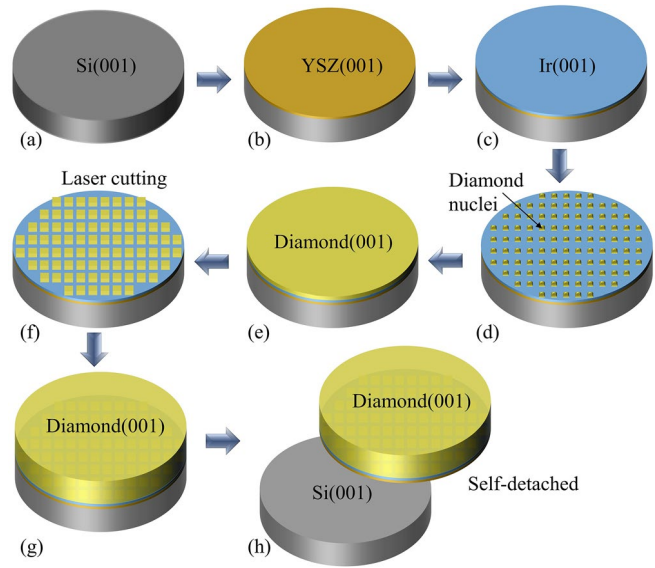

Figure 1. Schematic diagram of fabrication procedure of heteroepitaxial diamond on Ir/YSZ/Si (001). (a) 2-inch Si (001) substrate, (b)YSZ buffer layer deposition, (c)Ir buffer layer deposition, (d) diamond nucleation by BEN, (e) thin diamond growth, (f) laser patterned diamond template, (g) thick diamond growth, (h) diamond self-detachment.

Figure 1 illustrates schematically the fabrication process of the freestanding heteroepitaxial diamond grown on Ir/YSZ/Si (001) via laser-patterned templates presented in this paper. Firstly, a 100 nm thick YSZ layer was deposited on a 2-inch Si (001) single-crystal substrate using a PLD setup. An Ir epilayer with a thickness of approximately 120 nm was subsequently prepared through a magnetron sputtering apparatus. Details of preparing the YSZ and Ir buffer layers are described in our previous work.[24,25] After the preparation of Ir composite substrates, BEN treatment for diamond nucleation was performed on the prepared Ir/YSZ/Si (001) substrate using an MPCVD apparatus. The bias voltage was -300 V, and the duration was 40 mins. Onto the BEN-treated Ir composite substrate, a thin diamond film (D1) of around 50 nm was grown in the CVD chamber. A 355-nm UV laser with a power of 500 mW was employed to create 500 μm × 500 μm square patterns with lateral faces along <110> direction. The height and pitch of the diamond patterns were 50 nm and 20 μm, respectively. Following the laser-cutting procedure, the sample was subjected to an acid cleaning and an $H_2$ plasma etching treatment to remove contaminants and graphite residues left by the laser. The growth of a thick diamond layer (D2) was then carried out using the MPCVD reactor under optimized growth conditions. The diamond patterns underwent an epitaxial lateral overgrowth (ELO) process to achieve a coalescence, forming a closed and continuous diamond epilayer at the first 200 μm thickness. At the latter stage of the CVD growth, nitrogen was slightly added to the ambient gas to increase the growth rate. During the cooling process, the diamond/Ir layers automatically detached from the Si substrate due to the CTE-induced thermal strain and the incomplete attachment of the diamond to the Ir interface. The growth of the thick diamond lasted for 80 h, resulting in a final thickness of around 400 μm. The normal growth rate was 5 μm h$^{-1}$. Detailed parameters for BEN and CVD growth stages are summarized in Table 1.

Table 1. Parameters for BEN and diamond growth.

| Parameters | BEN | D1 | D2 |
| --- | --- | --- | --- |
| MW power (kW) | 1.5 | 5.5 | 5.5 |
| Pressure (Torr) | 23 | 50 | 50 |
| CH$_4$ in H$_2$ (%) | 7 | 2 | 4 |
| Bias voltage (V) | 300 | 0 | 0 |
| Temperature (°C) | 750 | 900 | 900 |
| Duration (h) | 0.67 | 0.25 | 80 |
| Thickness (μm) | 0 | 0.05 | 400 |

Figure 2 shows the X-ray diffraction (XRD) results of the heteroepitaxial diamond grown on Ir/YSZ/Si (001) for 80 h. The resultant diamond is around 400 μm thick. The XRD θ-2θ scan shown in Figure 2(a) exhibits four prominent diffraction peaks corresponding to Ir (200) (2θ=47.3°), Si (400) (2θ=69.21°), Ir (400) (2θ=106.7°), and diamond (400) (2θ=119.3°), verifying a pure (001)-oriented alignment and an excellent crystallinity. Notably, no diffraction peaks of the YSZ epilayer are detected since the thickness of the diamond crystal is much thicker than that of the YSZ layer.

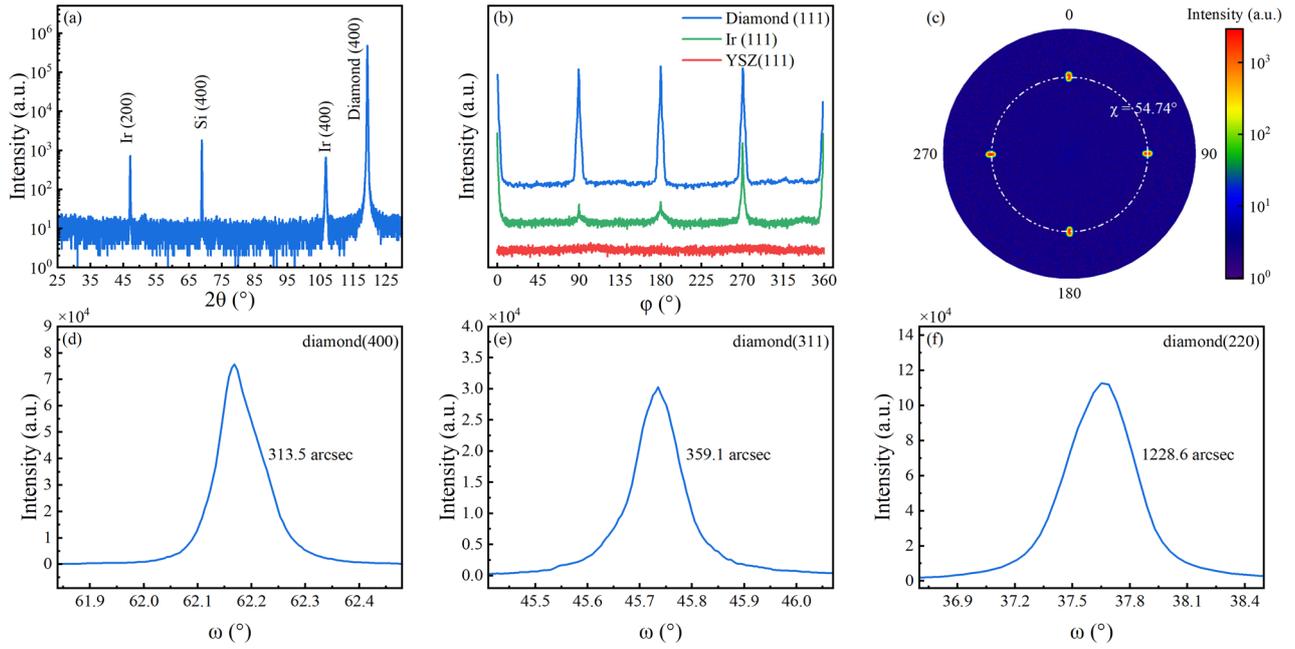

Figure 2. (a) X-ray diffractograms of θ-2θ scan of the heteroepitaxial diamond on Ir/YSZ/Si (001). (b) In-plane φ scan of diamond {111} and Ir (111) at a polar angle of χ=54.75°. (c) X-ray pole figure of diamond {111} diffraction peaks. XRCs of (d) diamond (400), (d) diamond (311), and (e) diamond (220) diffraction peaks from the freestanding diamond.

Mutual orientations among the diamond, Ir, YSZ, and Si substrate were verified by an in-plane φ scan of the {111} diffraction peaks of each layer at χ = 54.74°. As shown in Figure 2(b), the φ scan results of diamond and iridium layers show an exact quadruple symmetry with φ angles occurring at φ=0°, 90°, 180°, and 270°. No diffraction peaks of the YSZ layer and silicon substrate (not shown here) were detected, probably because they exceeded the penetration depth of X-rays. It was proved that the Ir/YSZ/Si substrate has a cube-on-cube crystallographic epitaxial relationship in our previous work. Therefore, this result demonstrates the same crystallographic epitaxial relationship in the diamond/Ir/YSZ/Si system. Furthermore, the X-ray pole figure of diamond {111} diffractions from the as-grown diamond crystal is shown in Figure 2(c). The absence of additional peaks except for four {111} diffraction peaks evidences that no twinning is found in the diamond crystal.

The XRCs of diamond (400), (311), and (220) diffractions are displayed in Figure 2(d)-(f), respectively. The FWHM value of the XRC for a specific crystalline facet can be used to characterize the epitaxial quality of the films. The FWHM of diamond (400) XRC corresponds to the azimuthal mosaic spread of diamond grains, and (311) relates to the polarization mosaic spread. The diamond (220) XRC enables evaluating the crystal quality along the surface. The FWHM of respective XRCs are 313.5, 359.1, and 1228.6 arcsecs, respectively, confirming the good epitaxial quality of the resultant diamond.

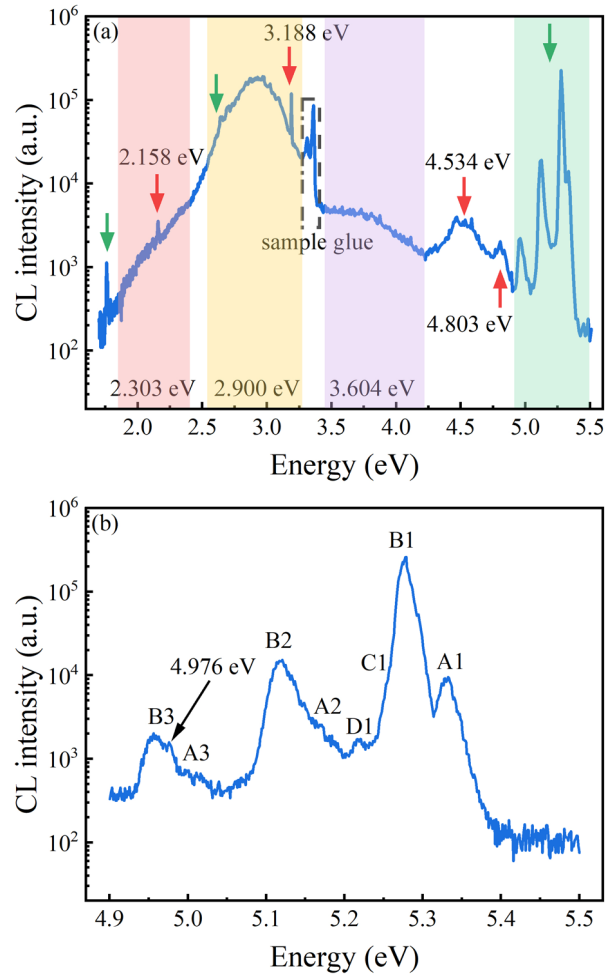

Figure 3. (a) CL spectrum of the diamond grown on Ir/YSZ/Si (001) measured at 12 K ranged from 220 nm to 720 nm, where the band-edge emissions and defect luminescence are marked with green and arrows respectively. (b) CL near-band-gap emission of the as-grown diamond.

The cathodoluminescence (CL) spectrum of the diamond measured at 12 K, ranging from 1.72 eV to 5.64 eV, is shown in Figure 3(a). Both band-edge emission (marked with green arrows) and defect luminescence (noted by red arrows) can be observed in the spectrum. Within the range of 4.8 eV to 5.5 eV, peaks resulting from near-band-gap emission are magnified and shown in Figure 3(b). The emission from the recombination of free excitons assisted by transverse acoustic (A1, 5.330 eV), transverse optic (B1, 5.277 eV), and longitudinal optical (C1, 5.255 eV) phonons can be clearly observed. Additionally, phonon replicas at the low-energy side of the CL spectrum are apparent due to multiple phonons emitted during free exciton recombination radiation. According to their peak locations, peaks A2 (5.172 eV) and A3 (5.016 eV) in Figure 3b are the phonons replica of peak A1; peaks B2 (5.117 eV) and B3 (4.958 eV) are the phonons replica of peak B1. Also, the lines at 2.626 eV and 1.765 eV represent the second and third-order diffraction peaks of the B2 line, respectively.[26] The well-defined near-band-gap emissions with a strong intensity agree with high crystalline quality.

Besides, the luminescence of impurities and defects is identified by several peaks and bands indicated by red arrows. The D1 line at 5.128 eV visible in Figure 3(b) is ascribed to boron-bound exciton emission with TO phonon. The appearance of two peaks, 4.976 eV in Figure 3(b) and 4.534 eV in Figure 3(a), is associated with the presence of boron.[27] Boron-related defects may originate from chamber contamination caused by our previously boron-doped diamond growth. Apart from boron-related defects, nitrogen-related spectral lines are apparent in Figure 3(a) due to the deliberate introduction of a small quantity of nitrogen gas during the latter growth stage to enhance the growth rate. The spectral line at 2.158 eV is attributed to the neutral state of nitrogen vacancy ($NV^0$).[28] The peak at 3.188 eV originates from defects formed by substitutional nitrogen atoms binding with interstitial carbon atoms.[29] The peak at 4.803 eV is attributed to excitons bound to lattice defects.[30] The spectral line ranging from 1.8 to 4.2 eV can be deconvoluted into three broad bands with central locations at 2.303 eV, 2.900 eV, and 3.604 eV. The broad band at 2.900 eV, designated as A-band, arises from closely spaced donor-acceptor (D-A) pairs, which are directly associated with the presence of dislocations.[31] Bands at 2.303 eV and 3.604 eV are associated with boron impurities.[30,32]

Figure 4 presents the results of Raman spectroscopy conducted on a 2 mm × 2 mm central area of a 400-micron-thick diamond crystal using an excitation wavelength of 532 nm. The analysis aims to evaluate pertinent information concerning the crystal quality and residual stress within the diamond. In Figure 4(a), the epitaxial diamond overlays the laser cutting patterns, and there is no significant difference in morphology between the diamond above the cutting stripes and that on the thin diamond. The distribution of laser stripes (shown darker) beneath can be observed through the diamond layer. Figure 4(b) is an optical microscope photograph of a portion of the test area illuminated with a 386 nm laser, which enables enhanced observation of the surface topography without interference from the laser stripes at the backside. Notably, the continuous nature of the diamond epitaxial layer is evident. Figure 4(c) shows the Raman spectrum with the first-order Raman peak of diamond located at 1332.02 cm$^{-1}$, which has a minimum FWHM value of 2.01 cm$^{-1}$ within the measured region after fitting with the Lorentz function. This peak width is close to the values of high-quality homoepitaxial single-crystal diamonds. Additionally, a weak and broad peak occurs at 1415 cm$^{-1}$ in the Stokes Raman spectrum, corresponding to the NV$^0$ center caused by introducing nitrogen during growth. Notably, the absence of a G-band peak indicative of graphite components near 1580 cm$^{-1}$ affirms the high purity of the diamond.

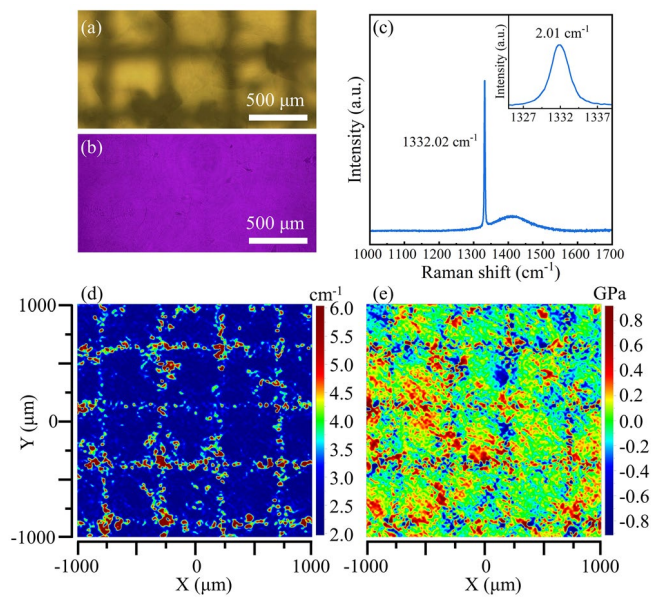

Figure 4. (a) An optical microscope photo of a part of the test area, using halogen illuminator. (b) An optical microscope photo of other parts of the test area illuminated with a 386 nm laser. (c) The diamond Raman spectrum with the minimum FWHM of 2.01 cm$^{-1}$ fitted within the test range. Inset exhibits the zoom-in graph of the diamond first-order Raman line. (d) The width distribution of the diamond first-order Raman line taken from the 2 mm × 2 mm area at the center of the sample. (e) A mapping image of residual stresses calculated based on the peak position shift of the diamond first-order Raman line.

Figure 4(d) illustrates the peak width distribution of the first-order Raman line of diamond within the test area. The measured peak widths at a series of test points range from 2.01 cm$^{-1}$ to 9.70 cm$^{-1}$, with an average value of 2.85 ± 0.36 cm$^{-1}$. These peak widths exhibit periodic variations corresponding to the laser patterns. Regions where laser ablation has removed the thin diamond layer predominantly exhibit higher peak widths (>3.5 cm$^{-1}$), while untreated regions consistently display lower values. Remarkably, the former widths are approximately two to three times wider than the latter. It is speculated that secondary nucleation, apart from epitaxial lateral growth, occurs after laser cutting, leading to compromised crystal quality in those areas subjected to laser treatment. Conversely, untreated regions maintain a favorable epitaxial relationship with the thin diamond layer and consequently demonstrate superior crystal quality throughout subsequent growth processes.

The evolution of stress during the growth process presents a complex phenomenon. Elevated overall stress is a significant factor impeding the growth of large thicknesses and sizes in heteroepitaxial diamonds. Employing the 1332.5 cm$^{-1}$ frequency as the center frequency of the Raman peak of stress-free diamonds, the mean stress within the test region was calculated, as shown in Figure 4(e). Stress levels range from -1.07 GPa to 1.39 GPa at test points, with an average biaxial in-plane stress of -0.03 ± 0.37 GPa. The stress distribution in the measured area shows a periodic pattern resembling that of the width distribution. Lager tensile/compressive stresses primarily appear in areas with a wider Raman line width. In contrast, the area without laser processing, especially the center of these areas with a lower FWHM, is almost stress-free. The presence of micro-strain, i.e., local deformation of the crystal lattice resulting from high concentrations of impurities or structural defects, brings about a broadening of the Raman line to some extent. This phenomenon mainly stems from additional defects introduced in laser-cut regions, leading to diminished crystal quality.

Figure 5(a) shows a picture of the 2-inch freestanding diamond crystal with a thickness of 400 μm. The dim color is a result of nitrogen impurities. Owing to the thermal strain and the partial attachment of the diamond to the Ir interface, the diamond/Ir layers detached from the Si substrate during the cooling process. This diamond plate was obtained without cracks and fracturing.

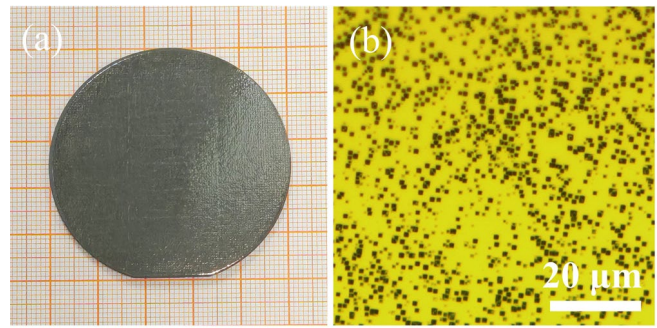

Figure 5. (a) Picture of the 2-inch freestanding diamond crystal grown on Ir/YSZ/Si (001). (b) Microscope image of the as-grown diamond surface after plasma etching to reveal dislocation.

To quantitatively evaluate dislocation density and its distribution, we subjected the diamond surface to $H_2/O_2$ plasma etching within the MPCVD setup. Under the etching conditions involving a microwave power of 3 kW, a gas pressure of 150 Torr, and a substrate temperature of 850°C, the process was executed for 20 minutes. This treatment aims to reveal dislocations appearing at the surface by inducing inverted pyramidal etch pits. The density of these dislocation features was determined to be approximately $2.2 \times 10^7$ cm$^{-2}$, a value consistent with the typical dislocation levels observed in heteroepitaxial diamonds.

In summary, a laser-patterned template method was developed during the heteroepitaxial growth of a 2-inch single-crystal diamond on an Ir/YSZ/Si (001) composite substrate. It is much simpler and less costly than the photolithography-patterned substrate. For the as-grown diamond crystal, the FWHMs of the XRC of the diamond (400) and (311) diffraction peaks are 313.5 and 359.3 arcsecs, respectively. The CL spectrum shows strong near-band gap-edge luminescence, confirming the good crystalline quality. Several spectral lines involving boron-related and nitrogen-related impurities and defects can still be visible due to chamber contamination and nitrogen incorporation during the growth process. The average peak width and residual stress calculated from Raman spectroscopy conducted on a 2 mm × 2 mm area of the diamond crystal are 2.85 ± 0.36 cm$^{-1}$ and -0.03 ± 0.37 GPa, respectively. The dislocation density determined by $H_2/O_2$ plasma etching is approximately 2.2×10$^7$ cm$^{-2}$. These results demonstrate the effectiveness of this method in preparing large-sized diamonds, which can also be used in other heteroepitaxial systems.